\documentclass[epj]{svjour}
\usepackage{graphicx}
\usepackage{amsmath}
\usepackage{bm}
\usepackage{amssymb}
\usepackage{latexsym}
\usepackage{dcolumn}
\usepackage{url}

\newcolumntype{d}{D{.}{.}{10}}
\begin{document}
\title{Analysis of X-ray spectra emitted from laser-produced plasmas of uranium}
\author{J.\ P.\ Marques\inst{1} \and
        P.\ Indelicato\inst{2} \and
        C. I. Szabo\inst{2}
        \and 
        F.\ Parente\inst{3}
%
}                     
\offprints{J.\ P.\ Marques}          
\institute{Centro de F{\'\i}sica At{\'o}mica e Departamento F{\'\i}sica, 
Faculdade de Ci{\^e}ncias, Universidade de Lisboa, \\
Campo Grande, Ed. C8, 1749-016 Lisboa, Portugal, 
\email{jmmarques@fc.ul.pt}
\and
Laboratoire Kastler Brossel,
\'Ecole Normale Sup\' erieure; CNRS; Universit\' e P. et M. Curie - Paris 6\\
Case 74; 4, place Jussieu, 75252 Paris CEDEX 05, France,
\email{paul.indelicato@spectro.jussieu.fr}
\and
Centro de F{\'\i}sica At{\'o}mica da Universidade de Lisboa e Departamento F{\'\i}sica da 
Faculdade de Ci{\^e}ncias e Tecnologia da Universidade Nova de Lisboa, Monte da Caparica, 2825-114 Caparica,  Portugal,
\email{facp@fct.unl.pt}
}
\date{Received: \today / Revised version: date}
%

\abstract{
In this paper, we used the multiconfiguration Dirac-Fock method to generate theoretical X-ray spectra for Co-, Ni-, Cu-, Zn-, Ga-, Ge-, As-, Se-, Br-, Kr-, and Rb-like uranium ions. Using the distribution of these ions in a  laser-produced plasma, for different plasma temperatures, we generate theoretical spectra which are compared to experimental data. 
\PACS{
      {32.30.Rj}{}   \and
      {31.15.A}{}   \and
      {32.70.Cs}{}
     } 
} 
\maketitle
%

\section{Introduction}
\label{intro}
    Hard X-ray spectra from laser-produced plasmas have \linebreak been the subject of extensive research in the last few years. This is mainly due to their importance in what concerns alternative energy sources and atomic, plasma and materials physics research, including astrophysics. In a recent review Hudson \textit{et al.} ~\cite{Hudson06} emphasize the importance of this subject for laser-driven inertial confinement fusion, extensively describing their challenges, the efforts that were under way in the world, and suggesting possible responses.
    
Usually, in this kind of plasmas a large number of atomic species are present, in different stages of ionization, and, within each of these stages, in different energy levels, the number of the latter being sometimes very large. Emitted X-radiation may be an important diagnostic of the plasma provided that one is able to connect the detected radiative lines with the emitting ion. Therefore, a calculation as precise as possible of the energy levels of the ions which may be present in a plasma is of crucial importance for the interpretation of recent high-resolution experimental spectra.

Seely \textit{et al.}~\cite{Seely03} reported on X-ray spectra from laser produced plasmas obtained when planar foils of U and Pb were bombarded with a total energy of about 12 kJ at the OMEGA (University of Rochester's Laboratory for Laser Energetics, LLE) laser facility. The spectra show a few intense and narrow features in the 12-22 keV energy range, interpreted as inner-shell transitions resulting from L-shell vacancies created by energetic electrons. The calculations of these authors suggest that the transitions are Ni-like or lower ionization stages.

Yuan and Moses~\cite{Yuan06} developed an atomic code, YAC, to calculate non-LTE plasma spectra using the detailed accounting method for line transitions.  With this code they obtained the ionic distributions for different plasma temperatures from 450 eV to 850 eV. Comparing their results with experimental Seely \textit{et al.}~\cite{Seely03} uranium spectra, they estimate that plasma temperature was 650 eV. The plasma conditions were estimated with the help of the 1D hydrodynamic code BUCKY~\cite{McFarlane95}.

In this work we use the Multi-Configuration Dirac-Fock code of Desclaux and Indelicato~\cite{Desclaux75,mcdf}, which includes QED corrections, to compute the energies of several X-ray lines from Co-like to Rb-like uranium ions which are important for the interpretation of  Seely \textit{et al.} uranium spectra.

Our results, together with the ionic distributions of Yuan and Moses allow for a more precise determination of the plasma temperatures in Seely \textit{et al.}~\cite{Seely03} experiment.

\section{Relativistic calculations}
\label{sec:relcalc}
We calculated bound-states wave functions using the 2007 version of the Dirac-Fock program of J. P. Desclaux and P. Indelicato~\cite{mcdf}. Details on the Hamiltonian and the processes used to build the wave-functions can be found elsewhere~\cite{Desclaux75,Desclaux93,Indelicato95,Indelicato96}.

The total wave function is calculated with the help of the variational principle. The total energy of the atomic system is the eigenvalue of the equation
\begin{equation}
\label{eq001}
        {\cal H}^{\mbox{{\tiny no pair}}}
        \Psi_{\Pi,J,M}(\ldots,\bm{r}_{i},\ldots)=E_{\Pi,J,M}
        \Psi_{\Pi,J,M}(\ldots,\bm{r}_{i},\ldots),
\end{equation}
where $\Pi$ is the parity, $J$ is the total angular momentum eigenvalue, and $M$ is the eigenvalue of its projection on the $z$ axis $J_{z}$. In this equation, the hamiltonian is given by
\begin{equation}
{\cal H}^{\mbox{{\tiny no pair}}} = \sum_{i=1}^{N} {\cal H}_D (r_i) + \sum_{i<j} V_{ij} (|\mathbf{r}_{ij}|),
\end{equation}
where ${\cal H}_D$ is the one electron Dirac operator and $V_{ij}$ is an operator representing the electron-electron of order one in $\alpha$. The expression of $V_{ij}$ in Coulomb gauge, and in atomic units, is
\begin{subequations}
\label{eq:eeinter}
\begin{align}
         V_{ij} =& \,\,\,\, \frac{1}{r_{ij}} \label{eq:coulop} \\
         &-\frac{\bm{\alpha}_{i} \cdot \bm{\alpha}_{j}}{r_{ij}} 
\label{eq:magop} \\ 
         & - \frac{\bm{\alpha}_{i} \cdot
         \bm{\alpha}_{j}}{r_{ij}} 
[\cos\left(\frac{\omega_{ij}r_{ij}}{c}\right)-1]
         \nonumber \\
        & + c^2(\bm{\alpha}_{i} \cdot
         \bm{\nabla}_{i}) (\bm{\alpha}_{j} \cdot
         \bm{\nabla}_{j})
         \frac{\cos\left(\frac{\omega_{ij}r_{ij}}{c}\right)-1}{\omega_{ij}^{2} 
r_{ij}},
         \label{eq:allbreit}
\end{align}
\end{subequations}
where $r_{ij}=\left|\bm{r}_{i}-\bm{r}_{j}\right|$ is the inter-electronic distance, $\omega_{ij}$ is the energy of the exchanged photon between the two electrons, $\bm{\alpha}_{i}$ are the Dirac matrices and $c$ is the speed of light. We use the Coulomb gauge as it has been demonstrated that it provides energies free from
spurious contributions at the ladder approximation level and must be used in many-body atomic structure calculations~\cite{Gorceix88,Lindroth89}.

The term (\ref{eq:coulop}) represents the Coulomb interaction, the term (\ref{eq:magop}) is the Gaunt (magnetic) interaction, and the last two terms (\ref{eq:allbreit}) stand for the retardation operator.
In this expression the $\bm{\nabla}$ operators act only on $r_{ij}$ and not on the following wave functions.

By a series expansion of the operators in expressions (\ref{eq:magop}) and (\ref{eq:allbreit}) in powers of $\omega_{ij}r_{ij}/c \ll 1$ one obtains the Breit interaction, which includes the leading retardation contribution of order $1/c^{2}$. The Breit interaction is, then, the sum of the Gaunt interaction (\ref{eq:magop}) and the Breit retardation
\begin{equation}
\label{eq:breit}
B^{\text{\scriptsize{R}}}_{ij} =
{\frac{\bm{\alpha}_i\cdot\bm{\alpha}_j}{2r_{ij}}} - 
\frac{\left(\bm{\alpha}_i\cdot\bm{r}_{ij}\right)\left(\bm{\alpha}_j
\cdot\bm{r}_{ij}\right)}{{2r_{ij}^3}}.
\end{equation}
In the many-body part of the calculation the electron-electron interaction is described by the sum of the Cou\-lomb and the Breit interactions. Higher orders in $1/c$, deriving from the difference
between Eqs.~(\ref{eq:allbreit}) and (\ref{eq:breit}) are treated here only as a first order perturbation. All calculations are done for finite nuclei using a Fermi distribution with a thickness parameter of 2.3 fm. The nuclear radii are taken from reference~\cite{Johnson85}.

The MCDF method is defined by the particular choice of a trial function to solve equation~(\ref{eq001}) as a linear combination of configuration state functions (CSF):
\begin{equation}
\left\vert \mathit{\Psi}_{\mathit{\Pi},J,M}\right\rangle =\sum_{\nu=1}%
^{n}c_{\nu}\left\vert \nu,\mathit{\Pi},J,M\right\rangle . \label{eq_cu}%
\end{equation}
The CSF are also eigenfunctions of the parity $\mathit{\Pi}$, the total angular momentum $J^{2}$ and its projection $J_{z}$. The label $\nu$ stands for all other numbers (principal quantum number, ...) necessary to define unambiguously the CSF. The $c_{\nu}$ are called the mixing coefficients and are obtained by diagonalization of the Hamiltonian matrix coming from the minimization of the energy in equation~(\ref{eq_cu}) with respect to the $c_{\nu}$.

The CSF are antisymmetric products of one-electron wave functions expressed as linear combination of Slater determinants of Dirac 4-spinors
\begin{equation}
\left\vert \nu,\mathit{\Pi},J,M\right\rangle =\sum_{i=1}^{N_{\nu}}%
d_{i}\left\vert
\begin{array}
[c]{ccc}%
\psi_{1}^{i}\left(  r_{1}\right)  & \cdots & \psi_{m}^{i}\left(  r_{1}\right)
\\
\vdots & \ddots & \vdots\\
\psi_{1}^{i}\left(  r_{m}\right)  & \cdots & \psi_{m}^{i}\left(  r_{m}\right)
\end{array}
\right\vert .
\end{equation}
where the $\psi$-s are the one-electron wave functions and the coefficients $d_{i}$ are determined by requiring that the CSF is an eigenstate of $J^{2}$ and $J_{z}$.
The $d_{i}$ coefficients are obtained by requiring that the CSF are eigenstates of $J^{2}$ and $J_{z}.$ The one-electron wavefunctions are defined as
\begin{equation}
\mathit{\psi} (r) = 
\left( 
\begin{array}{c}
\mathit{\chi}_{k}^{\mu} (\Omega) P(r) \\
i \mathit{\chi}_{-k}^{\mu} (\Omega) Q(r)
\end{array}
\right)
\end{equation}
where $\mathit{\chi}_{k}^{\mu}$ is a two-component spinor, and $P$ and $Q$ are respectively the large and small components of the wavefunction. A variational principle provides the integro-dif\-ferential equations to determine the radial wave functions and a Hamiltonian matrix that provides the mixing coefficients $c_{\nu}$ by diagonalization. One-electron radiative corrections (self-energy and vacuum polarization) are added afterwards. All the energies are calculated using the experimental nuclear charge distribution for the nucleus.

The so-called Optimized Levels (OL) method was used to determine the wave function and energy for each state involved. This method allows for a full relaxation of both initial and final states providing much better energies and wavefunctions. However, spin-orbitals in the initial and final states are not orthogonal, since they have been optimized separately. The formalism to take in account the wave functions non-orthogonality in the transition probabilities calculation has been described by L\"{o}wdin~\cite{Lowdin55}. The matrix element of a one-electron operator $O$ between two determinants belonging to the initial and final
states can be written
\begin{align}
&  \left\langle \nu\mathit{\Pi}JM\right\vert \sum_{i=1}^{N}O\left(
r_{i}\right)  \left\vert \nu^{\prime}\mathit{\Pi}^{\prime}J^{\prime}M^{\prime
}\right\rangle = \nonumber\\
& \times  \frac{1}{N!}   \left\vert
\begin{array}
[c]{ccc}%
\psi_{1}\left(  r_{1}\right)  & \cdots & \psi_{m}\left(  r_{1}\right) \\
\vdots & \ddots & \vdots\\
\psi_{1}\left(  r_{m}\right)  & \cdots & \psi_{m}\left(  r_{m}\right)
\end{array}
\right\vert \nonumber \\
& \times \sum_{i=1}^{m}O\left(  r_{i}\right)  \left\vert
\begin{array}
[c]{ccc}%
\phi_{1}\left(  r_{1}\right)  & \cdots & \phi_{m}\left(  r_{1}\right) \\
\vdots & \ddots & \vdots\\
\phi_{1}\left(  r_{m}\right)  & \cdots & \phi_{m}\left(  r_{m}\right)
\end{array}
\right\vert , \label{eq002}
\end{align}
where the $\psi_{i}$ belong to the initial state and the $\phi_{i}$ and primes belong to the final state. If $\psi=\left\vert n\kappa\mu\right\rangle $ and $\phi=\left\vert n^{\prime}\kappa^{\prime}\mu^{\prime}\right\rangle$  are orthogonal, i.e.,  $\left\langle n\kappa\mu|n^{\prime}\kappa^{\prime} \mu^{\prime}\right\rangle =\delta_{n,n^{\prime}}\delta_{\kappa,\kappa^{\prime }}\delta_{\mu,\mu^{\prime}}$,  the matrix element~(\ref{eq002}) reduces to one term  $\left\langle \psi_{i}\right\vert O\left\vert \phi_{i}\right\rangle$ where $i$ represents the only electron that does not have the same spin-orbital in the initial and final determinants. Since $O$ is a one-electron operator, only one spin-orbital can change, otherwise the matrix element is zero. In contrast, when the orthogonality between initial and final states is not enforced, one gets~\cite{Lowdin55}
\begin{equation}
\left\langle \nu\mathit{\Pi}JM\right\vert \sum_{i=1}^{N}O\left(  r_{i}\right)
\left\vert \nu^{\prime}\mathit{\Pi}^{\prime}J^{\prime}M^{\prime}\right\rangle
=\sum_{i,j^{\prime}}\left\langle \psi_{i}\right\vert O\left\vert
\phi_{j^{\prime}}\right\rangle D_{ij^{\prime}},
\end{equation}
where $D_{ij^{\prime}}$ is the minor determinant obtained by crossing out the $i$th row and $j^{\prime}$th column from the determinant of dimension $N\times N$, made of all possible overlaps $\left\langle \psi_{k}|\phi_{l^{\prime}}\right\rangle$.

Radiative corrections are also introduced, from a full QED treatment. The one-electron self-energy is evaluated using the one-electron values of Mohr and coworkers~\cite{Indelicato82,Mohr92,Mohr92a} and corrected for finite nuclear size~\cite{Mohr93}. The self-energy screening and vacuum polarization are treated with an approximate method developed by Indelicato and co-workers~\cite{Indelicato87,Indelicato90,Indelicato92,Indelicato98}.

\section{Results and discussion}
\label{sec:res}

In the analysis of their laser-produced uranium spectrum, using the HULLAC code~\cite{Shalom01}, Seely \textit{et al.}~\cite{Seely03} concluded that the observed energies of the two most intense features in the spectrum are 260 and 230 eV higher than the neutral U L$\alpha_1$ and L$\alpha_2$ energies, respectively. Other features were observed at energies 90 and 260 eV higher than the L$\beta_1$ energy. These energy differences led the authors to the conclusion that the observed transitions were due to highly charged ions in the vicinity of the Ni-like U or in lower ionization stages, around As- and Si-like U. 

Yuan and Moses~\cite{Yuan06} analysis of the same spectrum by means of an atomic model (YAC) which uses the detailed configurations as effective levels for the population equations but uses the detailed term accounting method for the lines transitions. Relativistic formulas are adopted in order to apply to high-Z elements. A comparison of the oscilattor strengths and energies between YAC and GRASP~\cite{Dyall89} results for Cu-like U 2p-4d transition shows that although the oscillator strengths agree very well, differences of 40 to 80 eV in the transition energies were found. Furthermore these authors present ionic distributions for different plasma temperatures from 450 to 850 eV and estimate, from their analysis, that the plasma is of a temperature of 650 eV.

In this work we calculated the energies of all levels in the L$_{i}$ ($i=1,2,3$)  M$_{i}$ ($i=1,\cdots,5$), N$_{i}$ ($i=1,\cdots,7$) and O$_{1}$ (for Rb only) one hole configurations in Co, -Ni-, Cu-, Zn-, Ga-, Ge-, As-, Se-, Br-, Kr- and Rb-like uranium. All transition energies and probabilities between initial L$_{i}$ and final M$_{i}$, N$_{i}$ and O$_{i}$ states were computed taking into account the  interaction of the inner holes with electrons in outer unfilled shells. In this calculation no electronic correlation was taken into account, but fully relaxed initial and final wavefunctions were independently calculated, so full relaxation in both energy and transition rates is included.

The intensity of the line corresponding to a transition from level $i$ to a level $j$ in a U$^{q+}$ ion with and in level $i$, is given by
\begin{equation}
\label{eq:I}
I_{ij}^{q} = h \nu_{ij} A_{ij}^q  N_i^{q}.
\end{equation}
In this equation $h \nu_{ij}$ and $A_{ij}^{q}$ are the radiative $i\rightarrow j$ transition energy and probability, respectively, and $N_i^{q}$ is the density of U$^{q+}$ ions in level $i$.  This density is obtained from the balance equations~\cite{Martins1} and equation (\ref{eq:I}) can be written as
\begin{eqnarray}
\label{eq:b}
I_{ij}^{q} &\propto &h \nu_{ij}   \frac{A_{ij}^q}{A_i^q}  \frac{N_i^{q}}{\sum_q N_i^q} \nonumber \\ 
&=& h \nu_{ij} \frac{A_{ij}^q}{\sum_j A_{ij} } \omega_i  \frac{N_i^{q}}{\sum_{q'} N_i^{q'}}.
\end{eqnarray} 
Here $A_i^q$ is the transition for deexcitation  by all possible processes (radiative and radiationless), $\omega_i$ is the fluorescence yield for level $i$ and the sum $\sum_{q'} N_i^{q'}$ is performed for all charged states $q'$. In this work the fluorescence yields given by Krause~\cite{Krause} for the neutral atoms were used.

In the case of Zn-like ($1s^2 2s^2 2p^6 3s^2 3p^6 3d^{10} 4s^2$) and Kr-like ($1s^2 2s^2 2p^6 3s^2 3p^6 3d^{10} 4s^2 4p^6$) ions we calculated the energies and transition probabilities for all lines. The results are given in table~\ref{tab:ZnKr}.

\begin{table*}[ht]
	\centering
	\caption{X-ray lines calculated energies (in eV) for Zn- and Kr-like uranium.}
		\begin{tabular}{cccccccc}
    & initial    	&     & final       &  Zn-like  &   Prob.    &  Kr-like  & Prob. \\
		\hline
L1	&	2s$_{1/2}$	&	M1	&	3s$_{1/2}$	&	16430.795	&	8.49E+10	&	16348.010	&	8.29E+10	\\
L1	&	2s$_{1/2}$	&	M2	&	3p$_{1/2}$	&	16782.763	&	1.38E+15	&	16702.777	&	1.35E+15	\\
L1	&	2s$_{1/2}$	&	M3	&	3p$_{3/2}$	&	17708.851	&	1.25E+15	&	17616.946	&	1.21E+15	\\
L1	&	2s$_{1/2}$	&	M4	&	3d$_{3/2}$	&	18234.176	&	6.62E+13	&	18162.897	&	6.49E+13	\\
L1	&	2s$_{1/2}$	&	M5	&	3d$_{5/2}$	&	18422.600	&	9.94E+13	&	18347.679	&	9.73E+13	\\
L1	&	2s$_{1/2}$	&	N1	&	4s$_{1/2}$	&	21095.510	&	5.31E+10	&	20866.003	&	4.95E+10	\\
L1	&	2s$_{1/2}$	&	N2	&	4p$_{1/2}$	&		&		&	21036.126	&	4.68E+14	\\
L1	&	2s$_{1/2}$	&	N3	&	4p$_{3/2}$	&		&		&	21340.323	&	5.14E+14	\\
L2	&	2p$_{1/2}$	&	M1	&	3s$_{1/2}$	&	15631.985	&	1.57E+14	&	15539.004	&	1.54E+14	\\
L2	&	2p$_{1/2}$	&	M2	&	3p$_{1/2}$	&	15983.953	&	2.19E+10	&	15893.772	&	2.12E+10	\\
L2	&	2p$_{1/2}$	&	M3	&	3p$_{3/2}$	&	16910.041	&	8.31E+12	&	16807.940	&	7.94E+12	\\
L2	&	2p$_{1/2}$	&	M4	&	3d$_{3/2}$	&	17435.366	&	5.47E+15	&	17353.891	&	5.36E+15	\\
L2	&	2p$_{1/2}$	&	M5	&	3d$_{5/2}$	&	17623.790	&	6.95E+11	&	17538.673	&	6.73E+11	\\
L2	&	2p$_{1/2}$	&	N1	&	4s$_{1/2}$	&	20296.700	&	5.29E+13	&	20056.998	&	5.12E+13	\\
L2	&	2p$_{1/2}$	&	N2	&	4p$_{1/2}$	&		&		&	20227.120	&	1.20E+10	\\
L2	&	2p$_{1/2}$	&	N3	&	4p$_{3/2}$	&		&		&	20531.317	&	3.27E+12	\\
L3	&	2p$_{3/2}$	&	M1	&	3s$_{1/2}$	&	11829.419	&	2.76E+14	&	11749.368	&	2.71E+14	\\
L3	&	2p$_{3/2}$	&	M2	&	3p$_{1/2}$	&	12181.388	&	3.59E+12	&	12104.136	&	3.48E+12	\\
L3	&	2p$_{3/2}$	&	M3	&	3p$_{3/2}$	&	13107.475	&	3.39E+12	&	13018.305	&	3.26E+12	\\
L3	&	2p$_{3/2}$	&	M4	&	3d$_{3/2}$	&	13632.801	&	4.59E+14	&	13564.256	&	4.50E+14	\\
L3	&	2p$_{3/2}$	&	M5	&	3d$_{5/2}$	&	13821.225	&	4.06E+15	&	13749.038	&	3.98E+15	\\
L3	&	2p$_{3/2}$	&	N1	&	4s$_{1/2}$	&	16494.135	&	8.75E+13	&	16267.362	&	8.41E+13	\\
L3	&	2p$_{3/2}$	&	N2	&	4p$_{1/2}$	&		&		&	16437.485	&	9.48E+11	\\
L3	&	2p$_{3/2}$	&	N3	&	4p$_{3/2}$	&		&		&	16741.682	&	1.17E+12	\\
	\hline
		\end{tabular}
\label{tab:ZnKr}
\end{table*}

For the other charge states we have to account for the interaction of the inner hole with the electrons in the unfilled outer shells. For example, for the 1s$^2$ 2s$^1$ 2p$^6$ 3s$^2$ 3p$^6$ 3d$^8$ 4s$^2$ configuration in Ni-like uranium with one hole in L$_{1}$ subshell, $J=1/2,\cdots,9/2$, leading to 20 energy levels. In the case of an hole in  L$_{2}$ or L$_{3}$ subshells, i.e., the 1s$^2$ 2s$^2$ 2p$^5$ 3s$^2$ 3p$^6$ 3d$^8$ 4s$^2$ configuration, $J$ values range from $1/2$ to $11/2$ leading to 45 levels. So, for the initial states we have a large number of levels, 65 in this case. In the final M$_{1}$ to N$_{1}$ states a similar situation will occur. The obvious consequence is that for a full calculation we have to compute a large number of transitions. In this case we calculated all transition probabilities between initial and final levels for $\Delta j=0,\pm1$ which represents 3234 lines. A similar situation occurs for all the other charge states that were calculated, except Zn-like and Kr-like uranium.The transition probabilities and energies calculated in this work can be found in~\cite{http}.

\begin{figure*}
	\centering
		\includegraphics{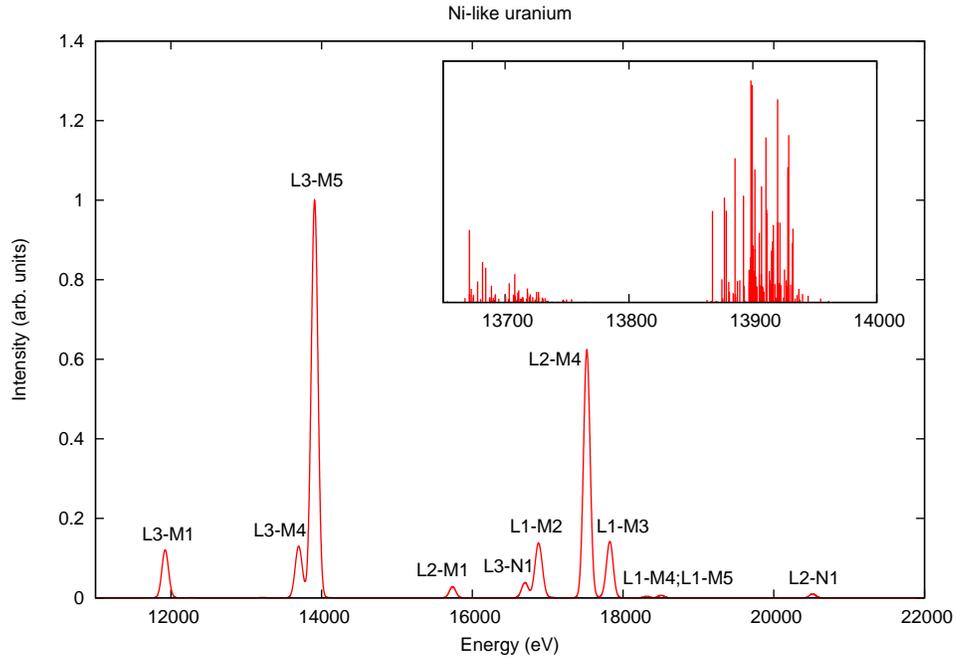}
	\caption{Calculated Ni-like uranium X-ray spectrum. This figure was generated using a Gaussian width of 95 eV. In the inset is presented a detailed view of the L$\alpha_{1,2}$ structure.}
	\label{fig:Ni}
\end{figure*}

In Fig.~\ref{fig:Ni} the theoretical spectrum obtained using for the lines a Gaussian distribution with a width of 95 eV, is presented for Ni-like uranium.  In the inset one can see a detailed view of the L$\alpha_{1,2}$ (L$_{3}-$M$_{4,5}$) structure.

\begin{figure*}
	\centering
		\includegraphics[width=13cm]{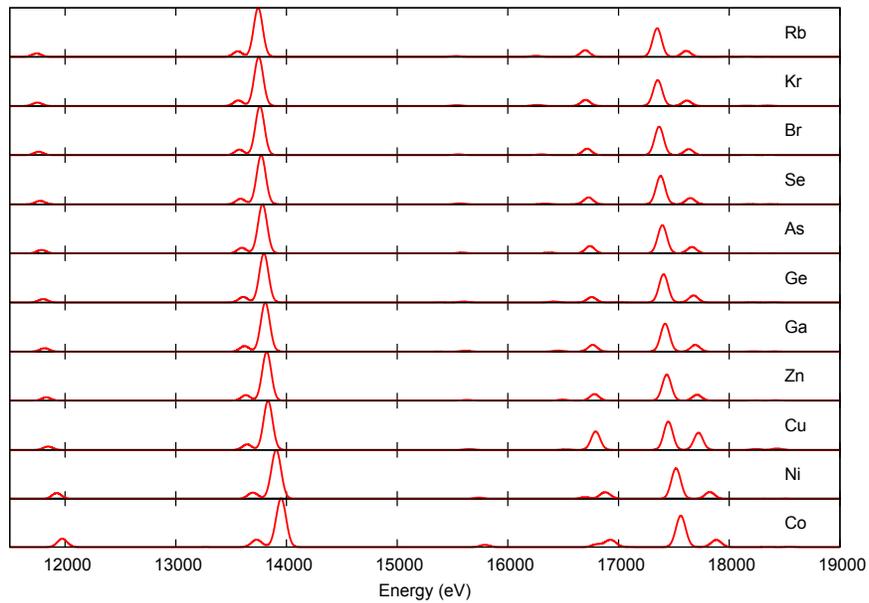}
	\caption{Calculated spectra for Co-, Ni-, Cu-, Zn-, Ga-, Ge-, As-, Se-, Br-, Kr-, and Rb-like uranium.}
	\label{fig:full_comp}
\end{figure*}

In Fig.~\ref{fig:full_comp} we compare the calculated spectra for Co-, Ni-, Cu-, Zn-, Ga-, Ge-, As-, Se-, Br-, Kr-, and Rb-like uranium. In this figure it is clear that as $Z$ increases, a shift of the peak energies to the lower energy side of the spectrum is observed.

\begin{figure*}[ht]
	\centering
		\includegraphics[width=13cm]{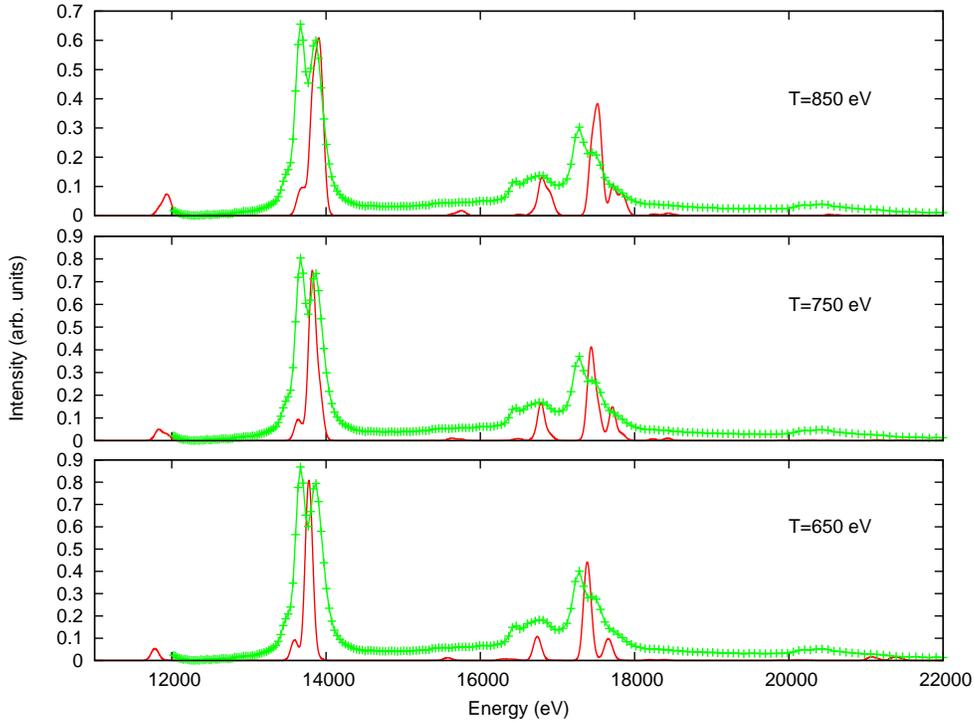}
	\caption{Comparison between experimental and calculates spectra for 650, 750, and 850 eV plasma temperatures (the intensity normalization was performed independently for each temperature).}
	\label{fig:comp_T}
\end{figure*}

In order to compare with the experimental spectrum obtained by Seely \textit{et al.}~\cite{Seely03} we
present in Fig.~\ref{fig:comp_T} the theoretical spectra calculated for plasma temperatures of 650, 750, and 850 eV, for a Gaussian width of 95 eV, plotted against the experimental data. For this purpose, we used the  ionic distributions for the different plasma temperatures, taken from Yuan and Moses~\cite{Yuan06}.

We conclude that the calculation for 850 eV plasma temperature is the one that best fits the experimental data, or, comparing the plots for 850 eV and 750 eV we may guess that the plasma temperature should be between these two values. Trying to verify this conclusion, we performed a $\chi^2$ test to see if the spectrum obtained with one of the temperatures matches better than the others the experimental data. The best matches were obtained for 750 and 850 eV which are close to each other, nevertheless the best match was obtained for a plasma temperature of 850 eV. This is in contrast with Yuan and Moses~\cite{Yuan06} conclusions that based in their own calculations pointed to a plasma temperature of 650 eV. 

As in the calculations by other authors the major disagreement with the experiment can be seen in the intensity ratio between the  L$\alpha_1$ and  L$\alpha_2$ lines. This behavior has been explained by opacity and plasma density gradient effects which may have an important role on the shape of the observed spectrum~\cite{Seely03}.

\section*{Acknowledgments}

We thank Doctor J. Seely for his kind collaboration in making available to us his X-ray spectrum for uranium.

This research was partially supported by the FCT pro\-ject POCTI/0303/2003 (Portugal), financed by the European Community Fund FEDER, and by the French-Portuguese collaboration (PESSOA Program, Contract n$^{\circ}$ 10721NF). Laboratoire Kastler Brossel is Unit{\'e} Mixte de Recherche du CNRS,  de l'{\'E}cole Normale Sup{\'e}rieure et de l'Universit{\'e} Pierre et Marie Curie, n$^{\circ}$ C8552.


\newpage

\end{document}